\documentclass[aps,prl,preprint,tightenlines,superscriptaddress,showpacs,byrevtex]{revtex4}
\usepackage{graphicx}
\usepackage{dcolumn}
\usepackage{bm}
\usepackage{amsmath}
\usepackage{amssymb}

\begin{document}

\preprint{\vbox{ \hbox{   }
                 \hbox{BELLE-CONF-0627}
                              }}
\title{\quad\\
[0.5cm] Search for $B^0 \to K^{*0} \nu \overline{\nu}$ Using One Fully
Reconstructed $B$ Meson}

\begin{abstract}
We present a search for the rare decay $B^0 \to K^{*0} \nu \overline{\nu}$,
using a data sample of 492~fb$^{-1}$ collected with the Belle
detector at the KEKB $e^+e^-$ collider. Signal candidates are required to
have an accompanying $B$ meson fully reconstructed in one of the hadronic
modes 
and signal-side particles consistent with a single $K^{*0}$ meson. 
No significant signal is observed in the data sample and we set a 90\%
confidence level upper limit of $3.4\times10^{-4}$ on the branching fraction.
\end{abstract}
\pacs{13.25.Hw, 14.40.Nd}


\affiliation{Budker Institute of Nuclear Physics, Novosibirsk}
\affiliation{Chiba University, Chiba}
\affiliation{Chonnam National University, Kwangju}
\affiliation{University of Cincinnati, Cincinnati, Ohio 45221}
\affiliation{University of Frankfurt, Frankfurt}
\affiliation{The Graduate University for Advanced Studies, Hayama} 
\affiliation{Gyeongsang National University, Chinju}
\affiliation{University of Hawaii, Honolulu, Hawaii 96822}
\affiliation{High Energy Accelerator Research Organization (KEK), Tsukuba}
\affiliation{Hiroshima Institute of Technology, Hiroshima}
\affiliation{University of Illinois at Urbana-Champaign, Urbana, Illinois 61801}
\affiliation{Institute of High Energy Physics, Chinese Academy of Sciences, Beijing}
\affiliation{Institute of High Energy Physics, Vienna}
\affiliation{Institute of High Energy Physics, Protvino}
\affiliation{Institute for Theoretical and Experimental Physics, Moscow}
\affiliation{J. Stefan Institute, Ljubljana}
\affiliation{Kanagawa University, Yokohama}
\affiliation{Korea University, Seoul}
\affiliation{Kyoto University, Kyoto}
\affiliation{Kyungpook National University, Taegu}
\affiliation{Swiss Federal Institute of Technology of Lausanne, EPFL, Lausanne}
\affiliation{University of Ljubljana, Ljubljana}
\affiliation{University of Maribor, Maribor}
\affiliation{University of Melbourne, Victoria}
\affiliation{Nagoya University, Nagoya}
\affiliation{Nara Women's University, Nara}
\affiliation{National Central University, Chung-li}
\affiliation{National United University, Miao Li}
\affiliation{Department of Physics, National Taiwan University, Taipei}
\affiliation{H. Niewodniczanski Institute of Nuclear Physics, Krakow}
\affiliation{Nippon Dental University, Niigata}
\affiliation{Niigata University, Niigata}
\affiliation{University of Nova Gorica, Nova Gorica}
\affiliation{Osaka City University, Osaka}
\affiliation{Osaka University, Osaka}
\affiliation{Panjab University, Chandigarh}
\affiliation{Peking University, Beijing}
\affiliation{University of Pittsburgh, Pittsburgh, Pennsylvania 15260}
\affiliation{Princeton University, Princeton, New Jersey 08544}
\affiliation{RIKEN BNL Research Center, Upton, New York 11973}
\affiliation{Saga University, Saga}
\affiliation{University of Science and Technology of China, Hefei}
\affiliation{Seoul National University, Seoul}
\affiliation{Shinshu University, Nagano}
\affiliation{Sungkyunkwan University, Suwon}
\affiliation{University of Sydney, Sydney NSW}
\affiliation{Tata Institute of Fundamental Research, Bombay}
\affiliation{Toho University, Funabashi}
\affiliation{Tohoku Gakuin University, Tagajo}
\affiliation{Tohoku University, Sendai}
\affiliation{Department of Physics, University of Tokyo, Tokyo}
\affiliation{Tokyo Institute of Technology, Tokyo}
\affiliation{Tokyo Metropolitan University, Tokyo}
\affiliation{Tokyo University of Agriculture and Technology, Tokyo}
\affiliation{Toyama National College of Maritime Technology, Toyama}
\affiliation{University of Tsukuba, Tsukuba}
\affiliation{Virginia Polytechnic Institute and State University, Blacksburg, Virginia 24061}
\affiliation{Yonsei University, Seoul}
  \author{K.~Abe}\affiliation{High Energy Accelerator Research Organization (KEK), Tsukuba} 
  \author{K.~Abe}\affiliation{Tohoku Gakuin University, Tagajo} 
  \author{I.~Adachi}\affiliation{High Energy Accelerator Research Organization (KEK), Tsukuba} 
  \author{H.~Aihara}\affiliation{Department of Physics, University of Tokyo, Tokyo} 
  \author{D.~Anipko}\affiliation{Budker Institute of Nuclear Physics, Novosibirsk} 
  \author{K.~Aoki}\affiliation{Nagoya University, Nagoya} 
  \author{T.~Arakawa}\affiliation{Niigata University, Niigata} 
  \author{K.~Arinstein}\affiliation{Budker Institute of Nuclear Physics, Novosibirsk} 
  \author{Y.~Asano}\affiliation{University of Tsukuba, Tsukuba} 
  \author{T.~Aso}\affiliation{Toyama National College of Maritime Technology, Toyama} 
% ...
\collaboration{The Belle Collaboration}


\collaboration{Belle Collaboration} \noaffiliation

\maketitle

\tighten

{\renewcommand{\thefootnote}{\fnsymbol{footnote}}} \setcounter{footnote}{0}

The decays $B \to K^{(*)} \nu \overline{\nu}$ and $B \to K^{(*)} l^+ l^-$
proceed through the flavor-changing neutral-current processes, which are
sensitive to 
physics beyond the Standard Model (SM) associated penguin loops. In the SM, 
the dominant diagram of these decays is the penguin process shown in Figure~\ref%
{fig:diagram} a). The SM branching fractions are estimated to be around the $%
10^{-5}/10^{-6}$ level for the $K^*/K$ modes based 
on next-to-leading-order calculations~\cite{ref:buchalla}. Calculation of
the decay amplitudes for $B \to K^{(*)} \nu \overline{\nu}$ is particularly
clean theoretically, owing to the absence of long-distance interactions that
affect the charged-lepton channels $B \to K^{(*)} l^+l^-$. New physics such 
as SUSY particles or the effect of a possible fourth generation could
potentially contribute to the penguin loop or box diagram (Figure~\ref%
{fig:diagram} b)) to enhance the branching fractions~\cite{ref:buchalla}.
Reference~\cite{ref:darkmatter} also discusses the possibility of
discovering light dark matter in $b\to s$ transitions with large
missing momentum.

\begin{figure}[htpb]
\begin{center}
\includegraphics[width=7cm]{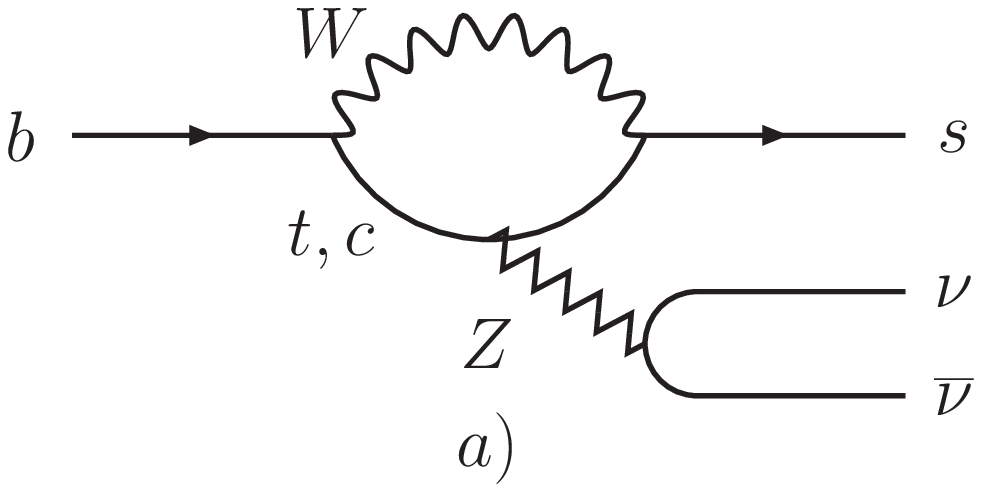} %
\includegraphics[width=7cm]{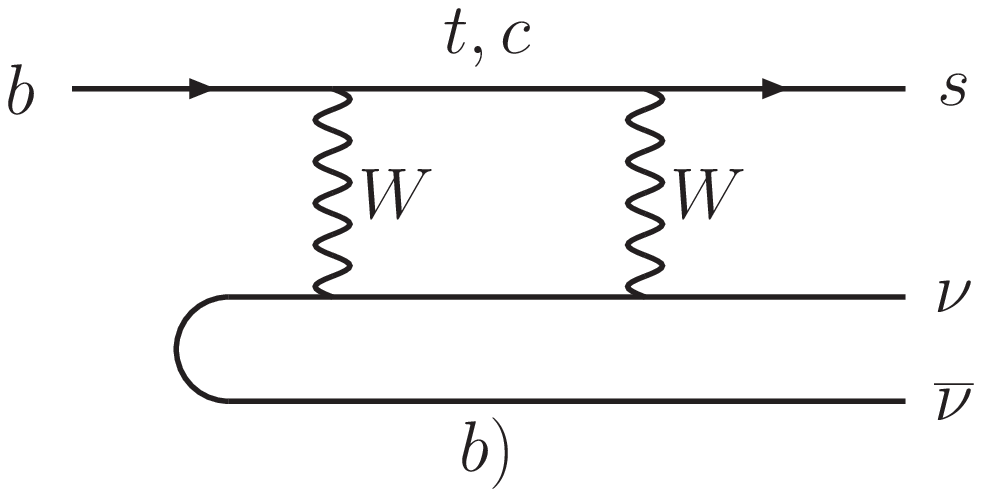}
\end{center}
\caption{The quark-level diagrams for $B \to K^{*} \protect\nu \overline{\protect\nu}$
decays.}
\label{fig:diagram}
\end{figure}

Experimental measurements~\cite{ref:kll} of the $b\to s$ transitions with
two charged leptons are in good agreement with SM calculations~\cite{ref:buchalla}. Further
investigation of the forward-backward asymmetry in 
$B\to K^* l^+l^-$~\cite{ref:afb} prefers the SM prediction although the
statistics are still limited. 
Due to the challenge of cleanly detecting rare modes with two final-state
neutrinos, only a few studies of $K^{(*)}\nu \overline{\nu}$ have been
carried out to date~\cite{Adam:1996ts}. 
In this paper, we report our first search for the decay $B^0 \to K^{*0} \nu 
\overline{\nu}$ using a data sample of 492 fb$^{-1}$ integrated luminosity
recorded at the $\Upsilon(4S)$ resonance, corresponding to $535\times 10^6$ $%
B$-meson pairs.

The Belle detector is a large-solid-angle magnetic spectrometer located at
the KEKB collider~\cite{ref:KEKB}, and consists of a silicon vertex detector
(SVD), a 50-layer central drift chamber (CDC), an array of 
aerogel threshold Cherenkov counters (ACC), a barrel-like arrangement of
time-of-flight scintillation counters (TOF), and an electromagnetic
calorimeter comprised of CsI(Tl) crystals (ECL) located inside a
superconducting solenoid that provides a 1.5~T magnetic field. An iron
flux-return located outside of the coil is instrumented to detect $K_L^0$
mesons and to identify muons (KLM). The detector is described in detail
elsewhere~\cite{ref:belle_detector}.

One of the $B$ mesons in the event is fully reconstructed as the tag-side $B$
candidate ($B_{\mathrm{tag}}$). The rest of the particles are assumed to be products of  
the signal-side $B$ meson ($B_{\mathrm{sig}}$). The $B_{\mathrm{tag}}$
candidates are reconstructed in one of the following decays: $B^0 \to
D^{(*)-} \pi^+$, $D^{(*)-}\rho^+$, $D^{(*)-}a_1^+$, and $D^{(*)-}D_s^{(*)+}$%
. The $D^-$ mesons are reconstructed as $D^- \to K^0_S\pi^-$, $%
K_S^0\pi^-\pi^0$, $K_S^0\pi^-\pi^+\pi^-$, $K^+\pi^-\pi^-$, and $%
K^+\pi^-\pi^-\pi^0$. The $D^{*-}$ mesons are reconstructed as $\overline
{D}{}
^0 \pi^-$, and the following decay channels are included for the $\overline{D
}{}^0$ mesons: $\overline{D}{}^0 \to K^{+}\pi^{-}$, $K^+\pi^-\pi^0$, $%
K^+\pi^-\pi^+\pi^-$, $K_S^0\pi^0$, $K_S^0\pi^-\pi^+$, $K_S^0\pi^-\pi^+\pi^0$
and $K^-K^+$. Furthermore, the $D_s^+\to K_S^0K^+$ and $K^+K^-\pi^+$ decays
are used for  $D_s^+$ mesons and $D_s^{*+} \to D_s^+ \gamma$
decay are also selected. A candidate $B_{\mathrm{tag}}$ meson is selected with the
beam constrained mass $M_{\mathrm{bc}} \equiv \sqrt{E_{\mathrm{beam}}^{2} -
p_{B}^{2}}$ and the energy difference $\Delta E \equiv E_{B} - E_{\mathrm{%
beam}}$, where $E_{B}$ and $p_{B}$ are the reconstructed energy and momentum
of the $B_{\mathrm{tag}}$ candidate in the $\Upsilon(4S)$ center-of-mass
(CM) frame, and $E_{\mathrm{beam}}$ is the beam energy in the same
system. The requirements on candidate $B_{\mathrm{tag}}$ mesons are $M_{\mathrm{bc}%
}>5.27$~GeV/$c^2$ and $-80$~MeV~$<\Delta E< 60$~MeV. If there are multiple 
$B_{\mathrm{tag}}$ candidates in one event, the candidate with the smallest $%
\chi^{2}$ based on the deviations from the nominal values of $\Delta E$, the 
$D$ meson mass, and the mass difference between the  $D^{*}$ and the $D$ (for the
candidate with a $D^*$ in the reconstruction) is chosen. 


Once a $B_{\mathrm{tag}}$ candidate is reconstructed, the remaining
particles are used to reconstruct a $B^{0}\rightarrow K^{\ast 0}\nu 
\overline{\nu }$ event. The $K^{\ast 0}$ candidate is reconstructed by two
charged tracks with opposite charge, and one of the tracks should have a
kaon likelihood greater than 0.6. The kaon likelihood is defined by $%
\mathcal{R}_{K}\equiv \mathcal{L}_{K}/(\mathcal{L}_{K}+\mathcal{L}_{\pi })$,
where $\mathcal{L}_{K}$ ($\mathcal{L}_{\pi }$) denotes a combined likelihood
measurement from the ACC, the TOF, and a $dE/dx$ from the CDC for the $K^{\pm }$ ($\pi ^{\pm }$%
) tracks. The daughter $K^{\pm }$ and $\pi ^{\mp }$ are required to have a
maximum distance to the interaction point (IP) of 2 cm in the beam direction
($z$) and of 5 cm in the transverse plane ($r$--$\phi $).  The invariant mass of
the $K^{\ast 0}$ candidate should be within a $\pm 75$ MeV$/c^{2}$ window of
the nominal $K^{\ast 0}$ mass. No other charged tracks or $\pi ^{0}$
candidates are allowed in the event, while a pair of photons with an
invariant mass within $\pm 18.5$~MeV$/c^{2}$ of the nominal $\pi ^{0}$
mass where each photon has an energy greater than $50$ MeV is considered a $%
\pi ^{0}$ candidate. A $B_{\mathrm{sig}}$ candidate is selected according to
the variable $E_{\mathrm{ECL}}\equiv E_{\mathrm{tot}}-E_{\mathrm{rec}}$,
where $E_{\mathrm{tot}}$ and $E_{\mathrm{rec}}$ are the total visible energy
measured by the ECL detector and the measured energy of reconstructed objects
including the $B_{\mathrm{tag}}$ and the signal side $K^{\ast 0}$ candidate,
respectively. A minimum threshold of 50 (100,150) MeV on the cluster energy
is applied for the barrel (forward endcap, backward endcap) region of the
ECL detector. The signal region is defined by $E_{\mathrm{ECL}}$~$<$~0.3~GeV
and the sideband region is given by 0.45~GeV~$<E_{\mathrm{ECL}}<$~1.2~GeV.

\begin{figure}[htpb]
\begin{center}
\includegraphics[width=8cm]{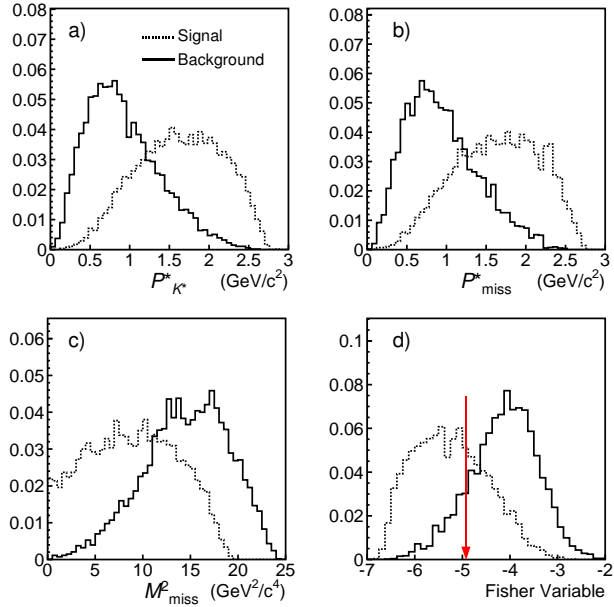}
\end{center}
\caption{The distributions for a) the momentum of $K^*$ candidate in the CM
frame ($P^*_{K^*}$), b) the missing momentum in the CM frame ($P^*_{\mathrm{%
miss}}$), c) the squared missing mass ($M^2_{\mathrm{miss}}$), and d) the
Fisher discriminant (arrow indicates the cut point). The dashed (solid) lines illustrate the distributions
from signal (background) MC samples.}
\label{fig:fisher_var}
\end{figure}

The dominant background source is generic $B^0\overline{B}{}^0$ decays. 
As shown in Figure~\ref{fig:fisher_var}, a Fisher discriminant containing three
input variables ($P^*_{K^*}$, $P^*_{\mathrm{miss}}$, and $M^2_{\mathrm{miss}}
$) is introduced to suppress the background, where $P^*_{K^*}$, $P^*_{%
\mathrm{miss}}$, and $M^2_{\mathrm{miss}}$ are the momentum of the $K^*$
candidate in the CM frame, the missing momentum in the CM frame, and the
squared missing mass. The missing momentum and missing mass are calculated 
using the momenta of the reconstructed $B_{\mathrm{tag}}$ and $K^{*0}$
candidate. Based on a figure of merit study, the events with a Fisher
discriminant value greater than $-4.9$ are rejected from the analysis.
Furthermore, the cosine of the angle between the missing momentum in the
laboratory frame and the beam pipe direction is required to be within the range  
$-0.86$ and $0.95$. These criteria suppress events with particles
produced along the beam pipe. 
The contributions from continuum background $e^+e^- \to q\overline{q%
}$~$(q=u,d,c,s)$ and other rare $B$ decays such as $B^0 \to K^{*0} \gamma$
are expected to be small. Based on Monte Carlo (MC) simulations, the
selection efficiency on the signal side is estimated to be $11\%$,
while the $B$ tagging efficiency is $0.087\%$. The reconstruction
efficiency in this analysis is equal to the product of the selection efficiency
and the $B$ tagging efficiency.

The contribution from each background source is examined with large MC
samples. The expected yields of background events in the signal region and
sideband region are $4.8\pm 1.5$ and $19\pm 3$ events, respectively. Further 
details are listed in Table~\ref{table:exp_yields}. From the full
experimental sample 13 events fall in the signal region and 
19 events fall in the sideband region. 

\begin{figure}[htpb]
\begin{center}
\includegraphics[width=7cm]{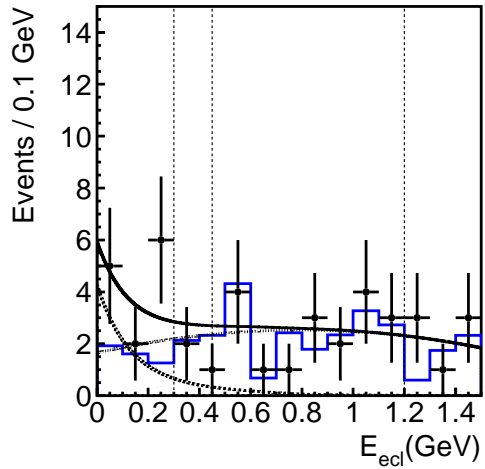}
\end{center}
\caption{The $E_{\mathrm{ECL}}$ distribution with fit results superimposed.
The solid histogram illustrates the background distributions from MC
simulations. The dashed, dotted and solid curves are the signal and
background components and their sum, respectively.}
\label{fig:Eecl_fit}
\end{figure}

\begin{figure}[htpb]
\begin{center}
\includegraphics[width=8cm]{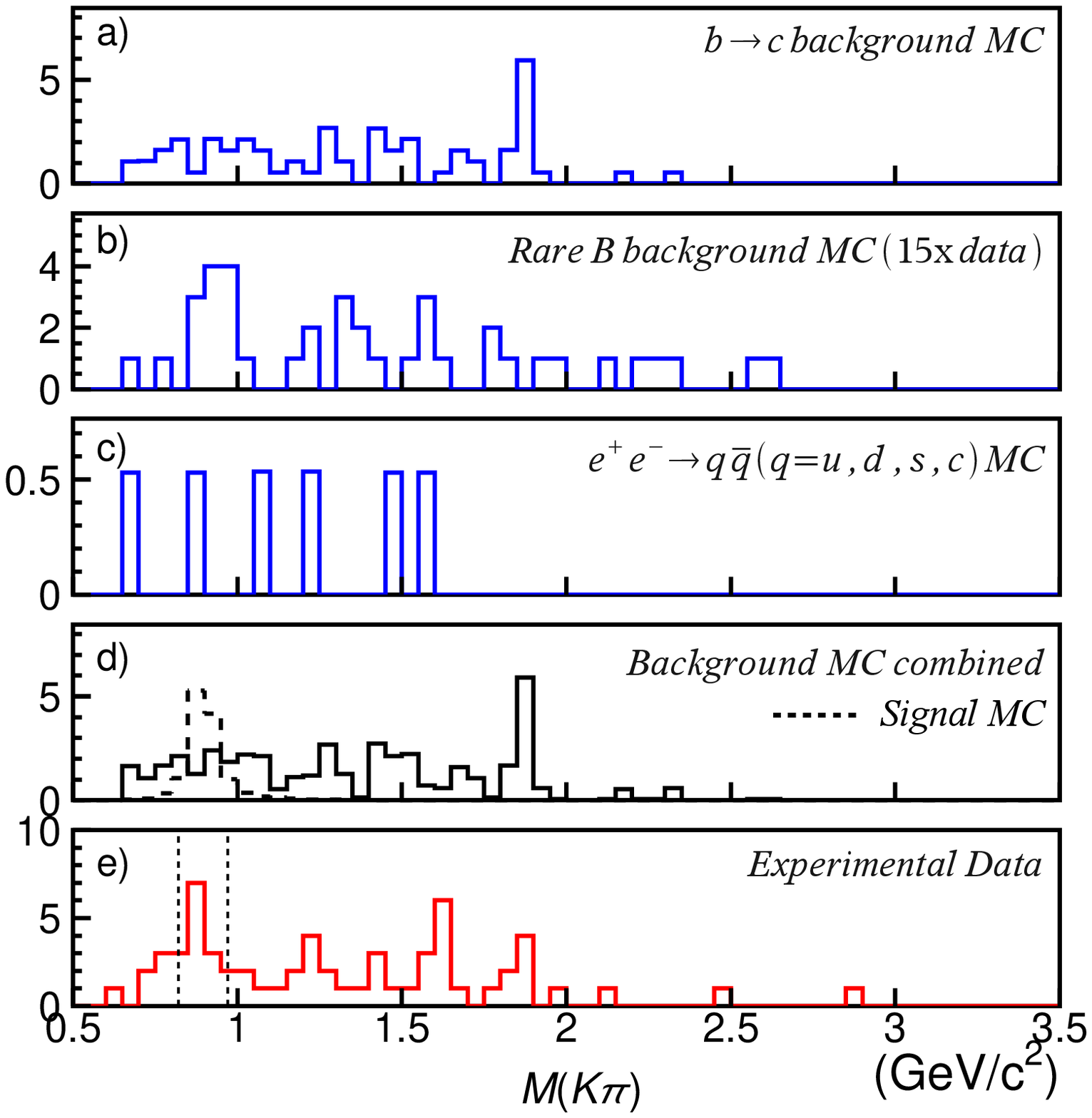}
\end{center}
\caption{The $M(K\protect\pi)$ distributions from the events in the signal
region, where a), b), c), d), and e) show the events from generic $b \to c$ MC,
rare $B$ MC, continuum $e^+e^-\to q\overline{q}$ $(q=u,d,c,s)$ MC,
background MC combined, and data, respectively. The dashed histogram
in d) illustrates the distribution for $B^0 \to K^* \protect\nu \overline{\protect\nu%
}$ signal MC.}
\label{fig:mkpi}
\end{figure}

\begin{figure}[htpb]
\begin{center}
\includegraphics[width=8cm]{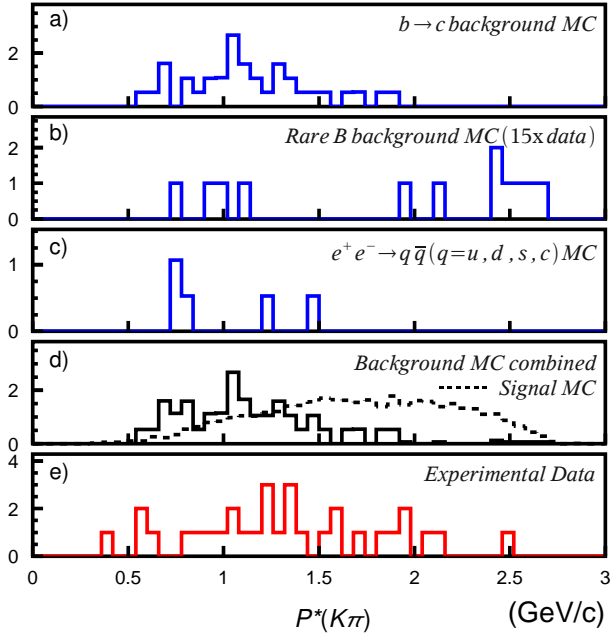}
\end{center}
\caption{The $P^*(K\protect\pi)$ distributions (without Fisher requirement)
from the events in the signal
region, where a), b), c), d), and e) show the events from generic $b \to c$ MC,
rare $B$ MC, continuum $e^+e^-\to q\overline{q}$ $(q=u,d,c,s)$ MC,
background MC combined, and data, respectively. The dashed histogram in
d) illustrates the distribution for $B^0 \to K^* \protect\nu \overline{\protect\nu}$
signal MC.}
\label{fig:pkst}
\end{figure}

The signal yield is extracted by a fit to
the $E_{\mathrm{ECL}}$ distribution. The signal probability density function
(PDF) is modeled by a smooth histogram obtained from a signal MC sample, and
a second order polynomial is used to describe the background
distribution. An extended likelihood function is introduced 
\begin{equation}
\mathcal{L}={\frac{{e^{-(N_{S}+N_{B})}}}{{N!}}}\prod_{i=1}^{N}\left[
N_{S}P_{S}(E_{\mathrm{ECL}}^{i})+N_{N}P_{B}(E_{\mathrm{ECL}}^{i})\right] ,
\end{equation}%
where $P_{S}$ and $P_{B}$ denote the signal and background PDF. The signal
yield, background yield, and total number of events in the fit are given by $%
N_{S}$, $N_{B}$, and $N$, respectively. We obtain a signal yield of $%
4.7_{-2.6}^{+3.1}$ events by maximizing the combined likelihood $\mathcal{L}$%
. The statistical significance is estimated to be 1.7$\sigma $ by a
comparison of the likelihood values for the best fit and a fit with zero
signal yield.
The $E_{\mathrm{ECL}}$ distribution with the fit results superimposed
is shown in Figure~\ref{fig:Eecl_fit}. The $M(K\pi )$ distributions for the
events in the signal region are also shown in Figure~\ref{fig:mkpi}, while the
momentum distributions of the reconstructed $K^{\ast 0}\rightarrow K\pi $
candidates are given in Figure~\ref{fig:pkst}.

\begin{table}[htpb]
\caption{The number of expected events in the signal and sideband regions.
}
\label{table:exp_yields}
\begin{center}
\begin{tabular}{lcc}
& Signal region & Sideband region \\ \hline
Continuum $e^+e^- \to q\overline{q} (q=u,d,c,s)$ & $0.5 \pm 0.5$ & $4.8 \pm
1.6$ \\ 
Generic $b \to c$ & $3.7 \pm 1.4$ & $13 \pm 3$ \\ 
Rare $B$ decays & $0.5 \pm 0.2$ & $1.1 \pm 0.3$ \\ \hline
Sum & $4.8 \pm 1.5$ & $19 \pm 3$ \\ \hline
Expected signal yield & $0.63\pm0.03$ & $0.10\pm0.01$ \\\hline
Data & 13 & 19 \\ 
&  & 
\end{tabular}%
\end{center}
\end{table}

The systematic uncertainty assigned for MC statistics is 3.8\%. By varying
the PDF parameters and repeating the fit and observing the variation of the results,
we infer an error associated with the fitting procedure of 5\%. The 
charged kaon (pion) identification is checked by a high statistics $D^*$
tagged sample, and an error of 0.8\% (1.2\%) is assigned. Other possible
systematic uncertainties such as tracking (2.1\%) and number of 
$B\overline{B}$ (1\%) are also included. The systematic uncertainty related to the $K^*$
mass selection is estimated to be 4\%. The total systematic uncertainty is
calculated to be 7.9\%. Considering the effects of both statistical and
systematic errors using an extension of the Feldman Cousins
method~\cite{Conrad:2002kn}, we obtain an upper limit of $\mathcal{B}(B^0 \to K^{*0} \nu 
\overline{\nu})<3.4\times 10^{-4}$ at the 90\% confidence level.

In conclusion, we have performed a search for the $B^0 \to K^{*0} \nu 
\overline{\nu}$ decays with a fully reconstructed $B$ tagging method on a
data sample of $535\times 10^{6}$ $B\bar{B}$ pairs collected at the $%
\Upsilon(4S)$ resonance with the Belle detector. We find $4.7^{+3.1}_{-2.6}$
signal events with a statistical significance of $1.7$ standard deviations.
The obtained upper limit is three times more stringent than 
the only published constraint~\cite{Adam:1996ts}.
Although the observed signal is much larger than the SM expectation, which is
estimated to be 0.63 events in the signal region, the error is large,
preventing us from drawing any conclusions at this point. We have examined
several signal-like candidates
and find that one of the events is consistent in missing
mass with a $B^0 \to K^{*0} \gamma$ decay. The hard photon in that event
hits the gap between the barrel and the forward endcap calorimeter.
Further understanding of the events in the signal region 
will require much larger $b \to c$ MC samples in addition
to more data. The limit on $B^0 \to K^{*0} \nu \overline{\nu}$ reported
here is still one order of magnitude above the prediction of Buchalla et
al.~\cite{ref:buchalla} and hence still allows room for substantial non-SM
contributions.


We thank the KEKB group for the excellent operation of the accelerator, the
KEK cryogenics group for the efficient operation of the solenoid, and the
KEK computer group and the National Institute of Informatics for valuable
computing and Super-SINET network support. We acknowledge support from the
Ministry of Education, Culture, Sports, Science, and Technology of Japan and
the Japan Society for the Promotion of Science; the Australian Research
Council and the Australian Department of Education, Science and Training;
the National Science Foundation of China and the Knowledge Innovation
Program of the Chinese Academy of Sciencies under contract No.~10575109 and
IHEP-U-503; the Department of Science and Technology of India; the BK21
program of the Ministry of Education of Korea, and the CHEP SRC program and
Basic Research program (grant No. R01-2005-000-10089-0) of the Korea Science
and Engineering Foundation; the Polish State Committee for Scientific
Research under contract No.~2P03B 01324; the Ministry of Science and
Technology of the Russian Federation; the Slovenian Research Agency; the
Swiss National Science Foundation; the National Science Council and the
Ministry of Education of Taiwan; and the U.S.\ Department of Energy.


\end{document}